\documentclass{PoS}
\usepackage[numbers,sort&compress]{natbib}

\title{
 Leptogenesis:\\ Improving predictions for experimental searches
}

\ShortTitle{Leptogenesis}

\author{\speaker{Marco Drewes}
\\
        Physik Department T70, Technische Universit\"at M\"unchen,
        James Franck Stra\ss e 1, D-85748 Garching, Germany \\
        E-mail: \email{marco.drewes@tum.de}}

\author{Bj\"orn~Garbrecht\\
        Physik Department T70, Technische Universit\"at M\"unchen,
        James Franck Stra\ss e 1, D-85748 Garching, Germany
        }
        
        \author{Dario Gueter\\
        Physik Department T70, Technische Universit\"at M\"unchen,
        James Franck Stra\ss e 1, D-85748 Garching, Germany\\        
        Max-Planck-Institut f\"ur Physik (Werner-Heisenberg-Institut), F\"ohringer Ring 6, 80805 M\"unchen, Germany\\
        Excellence Cluster Universe, Boltzmannstra{\ss}e 2, Technische Universit\"at M\"unchen, 85748 Garching, Germany
        }

\author{Juraj Klaric\\
        Physik Department T70, Technische Universit\"at M\"unchen,
        James Franck Stra\ss e 1, D-85748 Garching, Germany
        }
\abstract{Heavy right handed neutrinos could not only explain the observed neutrino masses via the seesaw mechanism, but also generate the baryon asymmetry of the universe via leptogenesis due to their CP-violating interactions in the early universe. We review recent progress in the theoretical description of this nonequilibrium process. Improved calculations are particularly important for a comparison with experimental data in testable scenarios with Majorana masses below the TeV scale, in which the heavy neutrinos can be found at the LHC, in the NA62 experiment, at T2K or in future experiments, including SHiP, DUNE and experiments at the FCC, ILC or CEPC. In addition, the relevant source of CP-violation may be experimentally accessible, and the heavy neutrinos can give a sizable contribution to neutrinoless double $\beta$ decay. In these low scale leptogenesis scenarios, the matter-antimatter asymmetry is generated at temperatures when the heavy neutrinos are relativistic, and thermal corrections to the transport equations in the early universe are large.}

\FullConference{38th International Conference on High Energy Physics\\
		3-10 August 2016\\
		Chicago, USA}

\begin{document}


\paragraph{Introduction} -
There is compelling evidence that the observable universe does not contain significant amounts of antimatter, 
and that the (baryonic) matter is the remnant of a tiny matter-antimatter asymmetry in the primordial plasma that survived after mutual annihilation of all particles and antiparticles, see e.g.\ Ref.~\cite{Canetti:2012zc} for a review.
This \emph{baryon asymmetry of the universe} (BAU) is commonly expressed in terms of the net baryon-to-photon ratio $\eta_B=n_B/n_\gamma\simeq 6\times 10^{-10}$. It cannot be explained within the Standard Model (SM) of particle physics and has to be generated dynamically via \emph{baryogenesis} if the radiation dominated epoch started with $\eta_B=0$, as suggested by many models of cosmic inflation. 
A particularly economic solution to the baryogenesis problem that relates $\eta_B$ to the observed neutrino masses and mixings
is given by \emph{leptogenesis} \cite{Fukugita:1986hr}. 
Consider the minimal extension of the SM Lagrangian $\mathcal{L}_{SM}$ by $n$ heavy Majorana neutrinos $N_i$,
\begin{eqnarray}
\label{eq:Lagrangian}
{\cal L}=\mathcal{L}_{\rm SM} + 
\frac{1}{2}\bar{N}_i({\rm i} \partial\!\!\!\!\!/-M_{ij}) N_j
-Y_{ia}^*\bar{\ell}_a \varepsilon\phi P_{\rm R} N_i
-Y_{ia}\bar{N}_iP_{\rm L}\phi^\dagger \varepsilon^\dagger \ell_a,
\end{eqnarray}
with $N_i=N_i^c$, where the superscript $c$ denotes a charge conjugation. 
The $N_i$ interact with the SM only via their Yukawa interactions $Y_{i a}$ with the SM lepton doublets $\ell_a$ ($a=e,\mu,\tau$) and the Higgs field $\phi$.  Here  $\varepsilon$ is the antisymmetric ${\rm SU}(2)$-invariant tensor with $\varepsilon^{12}=1$.
In the (type I) \emph{seesaw model} (\ref{eq:Lagrangian}), the same particles $N_i$ that generate the light neutrino mass matrix 
\begin{equation}\label{seesaw}
m_\nu = -v^2 Y^* M^{-1}Y^\dagger  = -\theta M \theta^T
\end{equation}
via the seesaw mechanism \cite{Minkowski:1977sc} 
 can produce a lepton asymmetry via their $CP$ violating interactions in the early universe, which is transferred into a BAU via electroweak sphaleron processes \cite{Kuzmin:1985mm}. 
 Here $\theta_{ai}=v Y^\dagger_{ai} M_i^{-1}$ (with $a=e,\mu,\tau$) are the active-sterile neutrino mixing angles at low energies, and $v=174$ GeV is the Higgs vev.
Various different authors have studied different implementations of this idea, which are e.g.\ summarised in the reviews \cite{Buchmuller:2005eh,Boyarsky:2009ix,Drewes:2013gca} 
and references therein. Here we only summarise recent developments in the field and give references for further reading. 

One can qualitatively distinguish two ways to generate the BAU in the minimal model (\ref{eq:Lagrangian}), either in the decay of the right handed neutrinos (\emph{freeze out scenario} or ``thermal leptogenesis'') \cite{Fukugita:1986hr}, or in $CP$ violating oscillations during their production (\emph{freeze in scenario} or ``baryogenesis from neutrino oscillations'') \cite{Akhmedov:1998qx,Asaka:2005pn}. Which one of them is realised depends on the masses $M_i$ of the $N_i$. For $M_i\gg v$, the right handed neutrinos come into equilibrium, freeze out and decay long before electroweak sphalerons freeze out at $T\simeq  130$ GeV. In this case the final asymmetry $\eta_B$ that can be observed today is usually created in the decay of the lightest right handed neutrino.
This scenario generically requires $M_i>10^9$ GeV \cite{Davidson:2002qv} unless the $N_i$ mass spectrum is highly degenerate \cite{Covi:1996wh}, leading to \emph{resonant leptogenesis} \cite{Pilaftsis:2003gt}.
For $M_i<v$, the seesaw relation (\ref{seesaw}) implies that at least some of the $Y_{ai}$ are much smaller than the electron Yukawa coupling. In this case a $\eta_B\neq0$ freezes in because the heavy neutrinos do not reach thermal equilibrium before sphaleron freezeout.

Leptogenesis is usually described in terms of momentum averaged semiclassical Boltzmann equations of the form 
\begin{eqnarray}
x H \frac{d Y_N}{d x}&=&-\Gamma_N(Y_N - Y_N^{\rm eq})\label{BE1}\\
x H \frac{d Y_{\rm B-L}}{d x}&=&(\epsilon \Gamma_N^{\rm D} + \epsilon' \Gamma_N^{\rm S})(Y_N - Y_N^{\rm eq}) - c_W \Gamma_N Y_{\rm B - L}.\label{BE2}
\end{eqnarray}
Here $Y_N$ is the abundance of heavy neutrinos and $Y_{\rm B-L}$ the total $B-L$ charge (both normalised to the cosmic entropy density),
$H$ is the Hubble rate, $\Gamma_N=\Gamma_N^{\rm D}+\Gamma_N^{\rm S}$ the thermal production rate of heavy neutrinos, including contributions from decays and inverse decays ($\Gamma_N^{\rm D}$) as well as scatterings ($\Gamma_N^{\rm S}$), and $c_W \Gamma_N$ is the washout rate. 
The parameters $\epsilon$ and $\epsilon'$ measure the difference between the rates for processes involving leptons and antileptons, normalised to their sum, and $c_W$ is a numerical factor that is governed by the ratios of the number densities of heavy neutrinos and SM leptons, which is of order unity if all particles are near thermal equilibrium.
We use $x=M/T$ as time variable, where $T$ is the temperature and $M$ an appropriately chosen mass scale (usually the mass of the lightest heavy neutrino). 
This description has been refined in various ways in recent years.

\paragraph{First principles derivation of kinetic equations} - The generation of a matter-antimatter asymmetry in leptogenesis is a pure quantum effect, as it relies on a CP violating quantum interference. It is not obvious that semiclassical Boltzmann equations, which are commonly used in cosmology, are suitable to describe this process in the dense primordial plasma quantitatively \cite{Buchmuller:2000nd}. This has motivated efforts to study leptogenesis from first principles of quantum field theory \cite{DeSimone:2007gkc,Anisimov:2010aq,Beneke:2010dz,
Frossard:2012pc}, in particular in the resonant case
\cite{DeSimone:2007edo,Garbrecht:2011aw,Garny:2009rv,Garny:2009qn,Garny:2011hg,Iso:2013lba,Dev:2014wsa} and in the context of flavour effects \cite{Beneke:2010wd,Dev:2014wsa} and in the freeze in scenario \cite{Drewes:2016gmt,Drewes:2016jae}. 
When all coupling constants are perturbative, then the first principles treatment at leading order reproduces the density matrix formalism \cite{Sigl:1992fn,Asaka:2005pn} commonly used in neutrino physics, assuming that the necessary resummations of all thermal corrections to the dispersion relations and rates are taken into account. The first principles approach provides a systematic way to 
calculate the coefficients in these equations 
that is inherently free of double counting problems, and in principle it allows to compute corrections to the leading order result (though this is practically challenging).  

\paragraph{Momentum averaging} - 
The effect of the momentum averaging has been studied systematically in Refs.~\cite{HahnWoernle:2009qn,Asaka:2011wq}. It was found that the averaged equations are accurate up to corrections between $\sim 10\%$ (for $M>T$) and order one (for $M<T$).  In the non-relativistic regime $M>T$, it has been shown that it is convenient to use the moments of the $N_i$ distribution function as dynamical variables  \cite{Bodeker:2013qaa} (instead of the occupation numbers for individual momentum modes).
Compared to tracking individual momentum modes, one can achieve the same accuracy with much less computational effort.

\paragraph{Quasiparticle dispersion relations} -
The dispersion relations of (quasi)particles in a hot plasma can considerably differ from those in vacuum. In the context of leptogenesis, this issue has been investigated in Ref.~\cite{Giudice:2003jh} and, in more detail, in Refs.~\cite{Kiessig:2011fw,Kiessig:2011ga,Beneke:2010dz,Garny:2011hg,Hohenegger:2014cpa,Miura:2013fxa,Dev:2014laa}.
In the freeze out scenario with strong washout, thermal corrections are small. 
In the weak washout regime with $T>M$ and in the freeze in scenario, where the BAU is generated at $T\gg M$, they have to be taken into account \cite{Asaka:2005pn,Shaposhnikov:2008pf,Canetti:2010aw,Canetti:2012vf,Canetti:2012kh,Drewes:2016lqo,Hernandez:2016kel,Drewes:2016gmt,Drewes:2016jae}. At tree level, the larger ``thermal mass'' of the Higgs boson implies that there exist 
three qualitatively different temperature regimes:
a low temperature regime where the decay $N\rightarrow \phi \ell$ is kinematically allowed, a high temperature regime where the decay $\phi\rightarrow N \ell$ is allowed, and an intermediate regime in which no decay is allowed \cite{Giudice:2003jh,Hambye:2016sby} and $\Gamma_N$ vanishes. However, a consistent resummation of all thermal contributions shows that the inclusion of scatterings lead to a smooth dependence of $\Gamma_N$ on $T$, and there is no regime with $\Gamma_N=0$ \cite{Anisimov:2010gy}\footnote{This effect may also play an important role in the context of perturbative cosmic reheating \cite{Drewes:2013iaa}.}, see next paragraph.  

\paragraph{Flavour effects} - If the couplings $Y_{ia}$ are of order unity, the seesaw  relation (\ref{seesaw}) suggests $M_i>10^{14}$ GeV, and the BAU is generated in the decay of $N_i$ at temperatures $T>10^{12}$ GeV (freezeout scenario), at which the SM flavours are indistinguishable ("vanilla leptogenesis"). Then $\eta_B$ is independent of the phases in the light neutrino mixing matrix. 
For smaller values of $M_i$ and $T<10^{12}$ GeV, the charged lepton Yukawa couplings affect the evolution of leptonic asymmetries, and flavour effects are relevant \cite{Barbieri:1999ma,Blanchet:2006be,Abada:2006ea,Abada:2006fw,Nardi:2006fx,JosseMichaux:2007zj}. 
In this case (\ref{BE1},\ref{BE2}) should be replaced by matrix valued generalisations \cite{Sigl:1992fn}, which have e.g.\ been derived in Refs.~\cite{Asaka:2005pn,Cirigliano:2009yt,Beneke:2010wd,Fidler:2011yq,Canetti:2012kh,Dev:2014wsa,Dev:2014laa,Drewes:2016gmt}.
Flavour effects introduce a dependence of $\eta_B$ on the phases in the light neutrino mixing matrix and reduce the lower bound on non-degenerate $M_i$ for leptogenesis in the freezeout scenario to $M_i>10^6$ GeV \cite{Antusch:2009gn,Racker:2012vw}.
The intermediate regime between the vanilla scenario and the fully flavoured scenario has been studied in detail in Ref.~\cite{Beneke:2010dz}. If one wants to achieve successful leptogenesis with even smaller $M_i$, this requires a mass degeneracy to ensure a resonant enhancement of $\eta_B$. Moreover, sizeable $Y_{ia}$ can only be made consistent with small neutrino masses if there are cancellations in (\ref{seesaw}).
Both can be achieved naturally in models with an approximate conservation of $B-L$, where leptogenesis relies mostly on flavour effects \cite{AristizabalSierra:2009bh,Racker:2012vw}.
In the freeze in scenario, the asymmetry is always generated at $T\gg M_i$, where the total lepton number is negligible \cite{Akhmedov:1998qx,Asaka:2005pn}, and baryogenesis relies entirely on flavour effects.

\paragraph{Thermal production and damping rates} - The rate $\Gamma_N$ at $T<M$ is in good approximation given by the $N_i$ vacuum decay rate \cite{Salvio:2011sf,Laine:2011pq,Biondini:2013xua,Anisimov:2010gy,Garbrecht:2013urw,Ghisoiu:2014ena}. For $T>M$, on the other hand, thermal effects can dominate. These include quantum statistical factors, modifications of the (quasi)particle dispersion relations in the plasma and the fact that the $N_i$ can be produced in scatterings in the dense plasma. For $T\gg M$, there is a collinear enhancement of multiple scatterings, and the Landau-Pomeranchuk-Migdal effect should be taken into account \cite{Besak:2010fb,Anisimov:2010gy,Besak:2012qm,Ghisoiu:2014ena}. The dominant contribution to $\Gamma_N$ in this regime comes from logarithmically enhanced t-channel scatterings \cite{Garbrecht:2013gd,Garbrecht:2013urw}.\footnote{These calculations have been performed in the minimal model (\ref{eq:Lagrangian}). In theories with an extended Higgs sector, there can be additional contributions from decays of the heavy scalars, see e.g. \cite{Drewes:2015eoa} and references therein.} 
In Ref.~\cite{Ghiglieri:2015jua,Ghiglieri:2016xye}, this calculation has been extended to temperatures across and below the electroweak scale. Thermal corrections do not only affect $\Gamma_N$, but also the washout rate $c_W \Gamma_N$ \cite{Bodeker:2014hqa,Ghisoiu:2014ena} and its flavoured generalisations \cite{Garbrecht:2014aga,Garbrecht:2013urw,Ghiglieri:2016xye}.
In the freeze in scenario, a consistent analysis also requires to include the temperature dependence of the sphaleron rate across the electroweak crossover \cite{D'Onofrio:2012jk,D'Onofrio:2014kta}.

\paragraph{Spectator effects} - The effect of spectator processes \cite{Buchmuller:2001sr} that do not change the total asymmetry, but reshuffle it between different degrees of freedom, has recently been revisited in both, the 
freeze out \cite{Garbrecht:2014kda} and freeze in \cite{Shuve:2014zua,Hernandez:2016kel,Drewes:2016gmt,Drewes:2016jae} scenarios. In the latter the spectators ca have a significant effect on the viable parameter space.

\paragraph{$CP$ violating parameter} - Thermal corrections to the $CP$ violating parameter $\epsilon$ have been studied by several authors \cite{Garny:2010nj,Beneke:2010wd,Garny:2011hg,Garbrecht:2011aw,Dev:2014laa,Dev:2014wsa,Biondini:2015gyw,Biondini:2016arl}. A question of particular interest is the dependence of $\eta_B$ on the $N_i$ mass spectrum in the regime where the splitting $\Delta M$ of two $M_i$ is much smaller than their average mass $\bar{M}$, leading to a resonant enhancement that allows to generate the observed $\eta_B$ with $\bar{M}$ below the TeV scale \cite{Pilaftsis:2003gt}. In the resonant limit $\Delta M/\bar{M}\rightarrow 0$, $\eta_B$ is regularised by the finite thermal width $\Gamma_N$. The precise form of the "regulator" that determines the maximal $\eta_B$ has been the subject of a long standing dispute \cite{Liu:1993ds,Covi:1996fm,Flanz:1996fb,Pilaftsis:1997dr,Buchmuller:1997yu,Pilaftsis:2003gt}. The issue has been studied from first principles in Refs.~\cite{Garny:2011hg,Dev:2014laa,Dev:2014wsa,Garbrecht:2011aw,Dev:2014wsa,Kartavtsev:2015vto}, but disagreement on how to count (and not double count) contributions from "mixing" and "oscillations" of the $N_i$ remains, leading to factors $\sim2$ uncertainty in these contributions.
Note that, while parameter relevant for the enhancement of the source for the total lepton asymmetry is $\bar{M}^2/\Delta M^2$, the flavoured source is enhanced by $T^2/\Delta M^2$,
which allows to generate the observed $\eta_B$ with $M_i$ below the electroweak scale without a mass degeneracy \cite{Drewes:2012ma}.
The $CP$ violating phases in the $Y_{ia}$ can be experimentally constrained from light neutrino oscillations at DUNE and NOvA, 
measurements of the mixings $U_{ai}^2=|\theta_{ai}|^2$ in direct search experiments \cite{Shaposhnikov:2008pf,Asaka:2011pb,Hernandez:2016kel,Drewes:2016jae}
and in the $CP$ violation in $N_i$ decays \cite{Cvetic:2014nla}.\footnote{See e.g. Ref.~\cite{Drewes:2016blc} and references therein for a recent discussion of the relation between $\eta_B$ and observable $CP$ violation in the lepton sector.}

\begin{figure}
	\centering
	\includegraphics[width=0.45\textwidth]{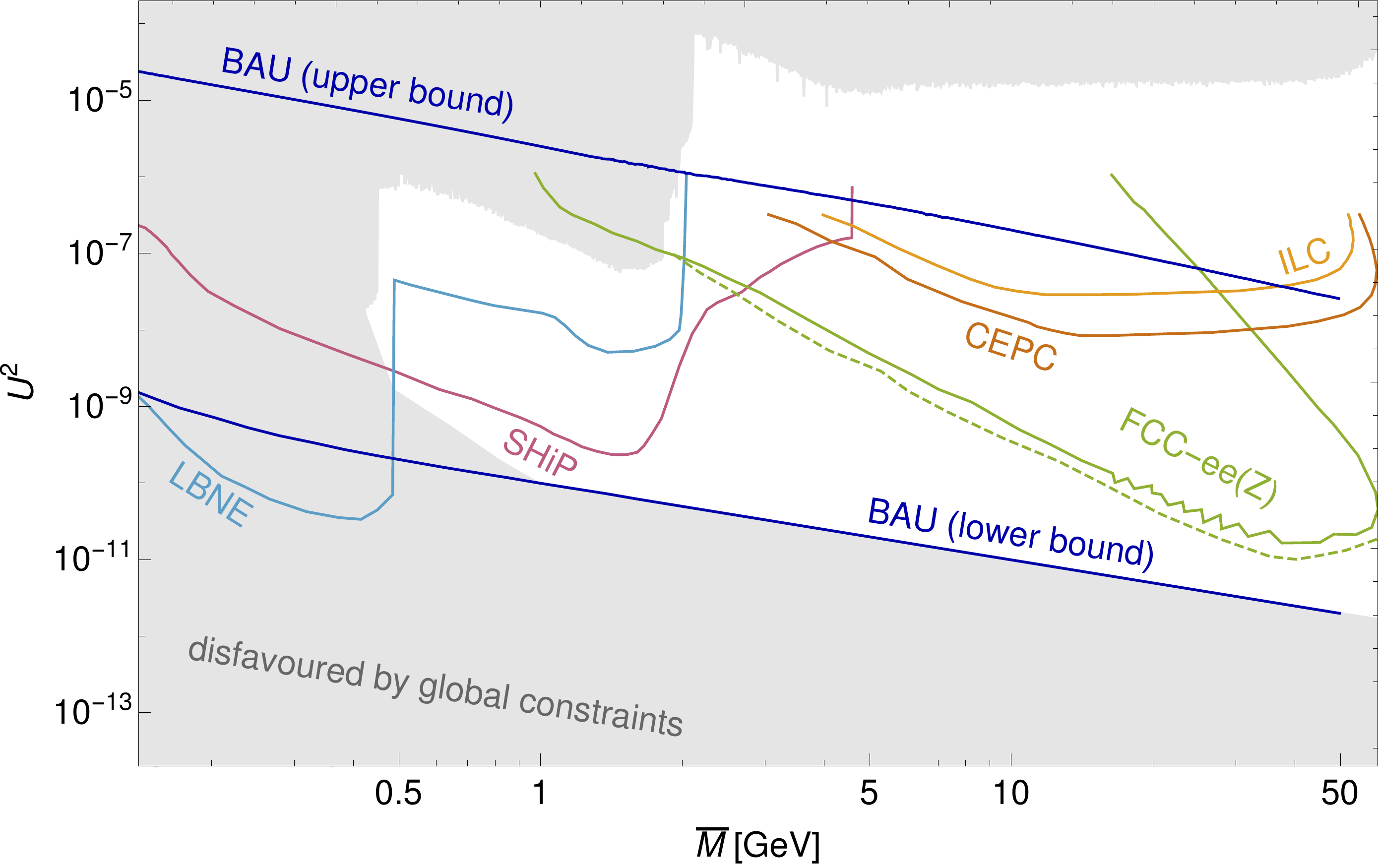}
	\includegraphics[width=0.45\textwidth]{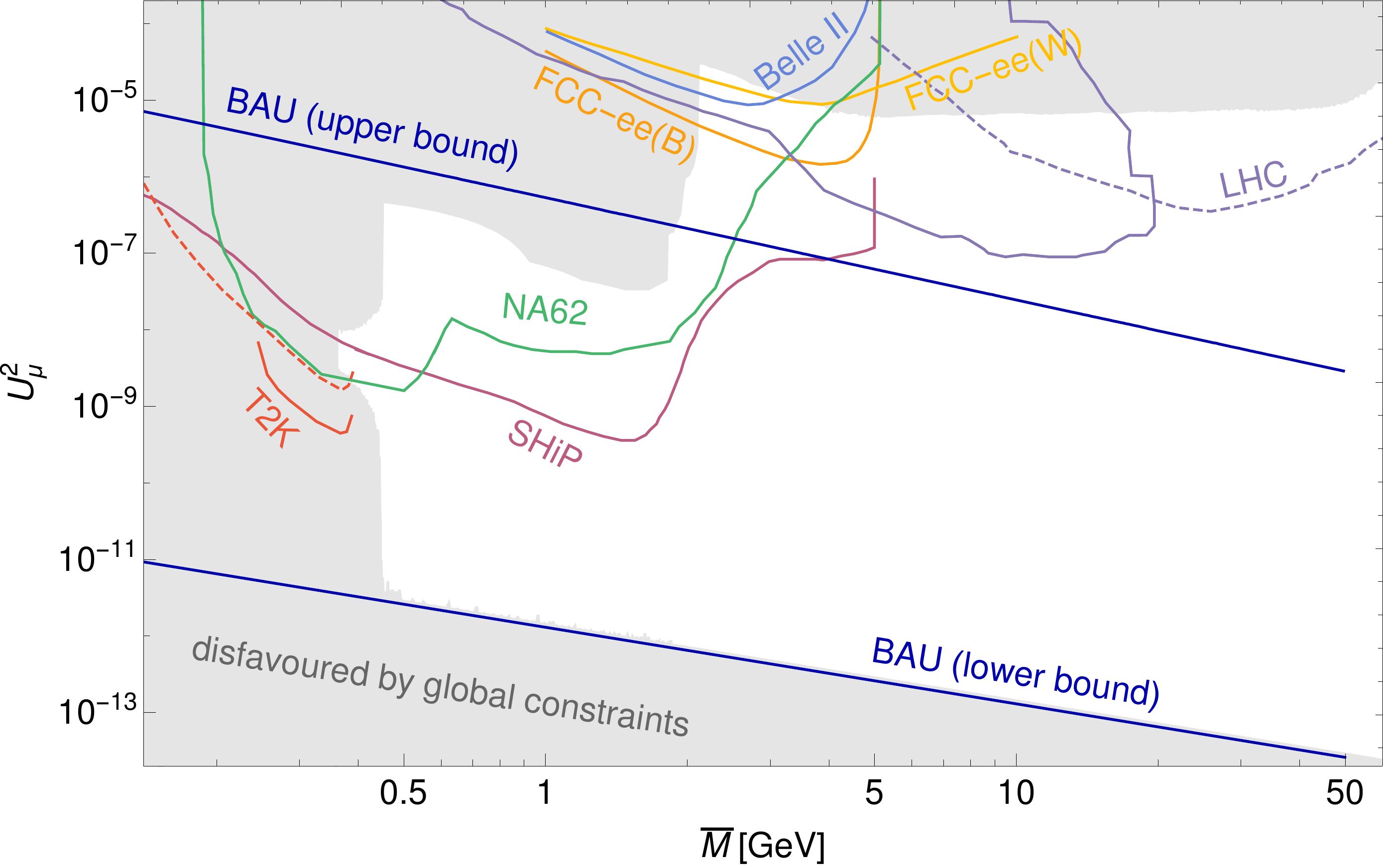}
\caption{
Viable leptogenesis parameter space for inverted hierarchy and $n=2$ (between the blue lines) compared to the constraints from past experiments (grey area) and the reach of some future experiments indicated in the plot, as found in Ref.~\cite{Drewes:2016jae}. Here $U^2=\sum_a U_a^2$.
\label{plot}
}
\end{figure}
\begin{figure}
	\includegraphics[width=0.45\textwidth]{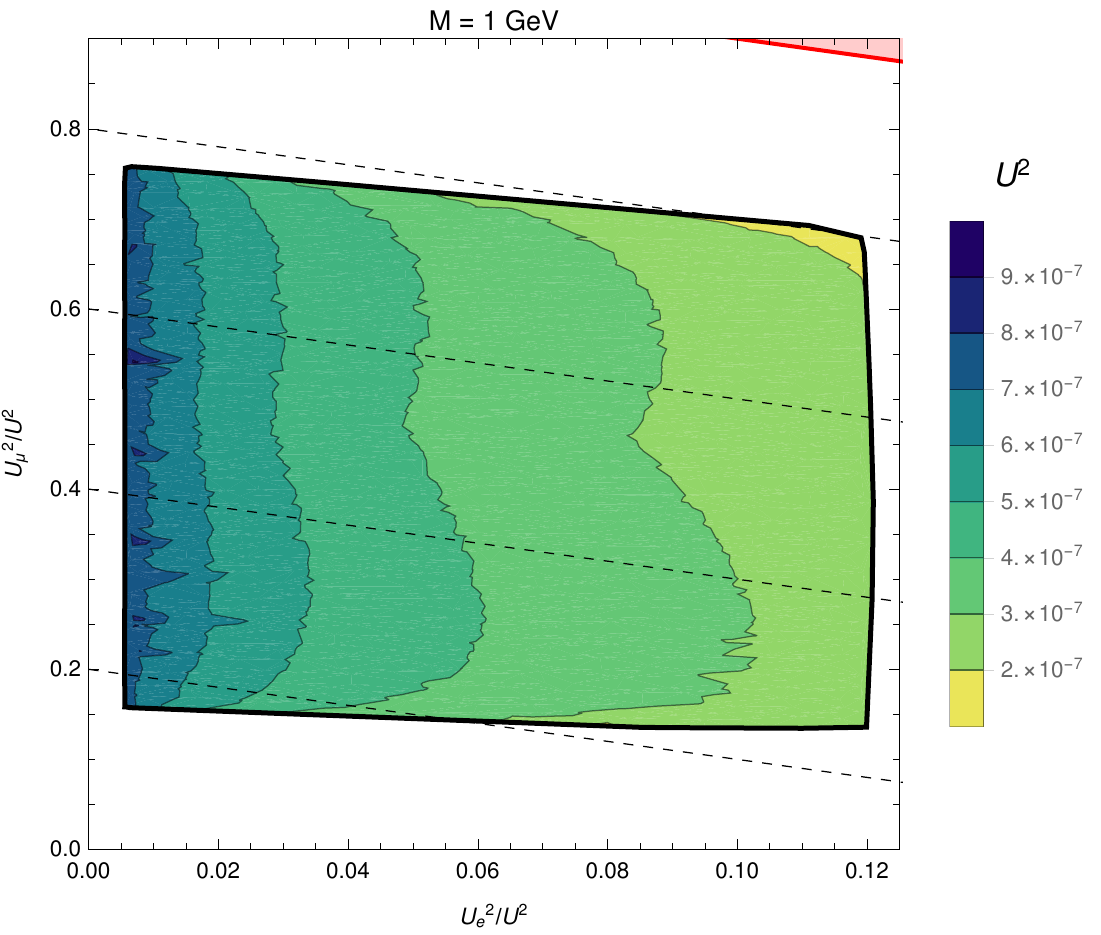}\includegraphics[width=0.45\textwidth]{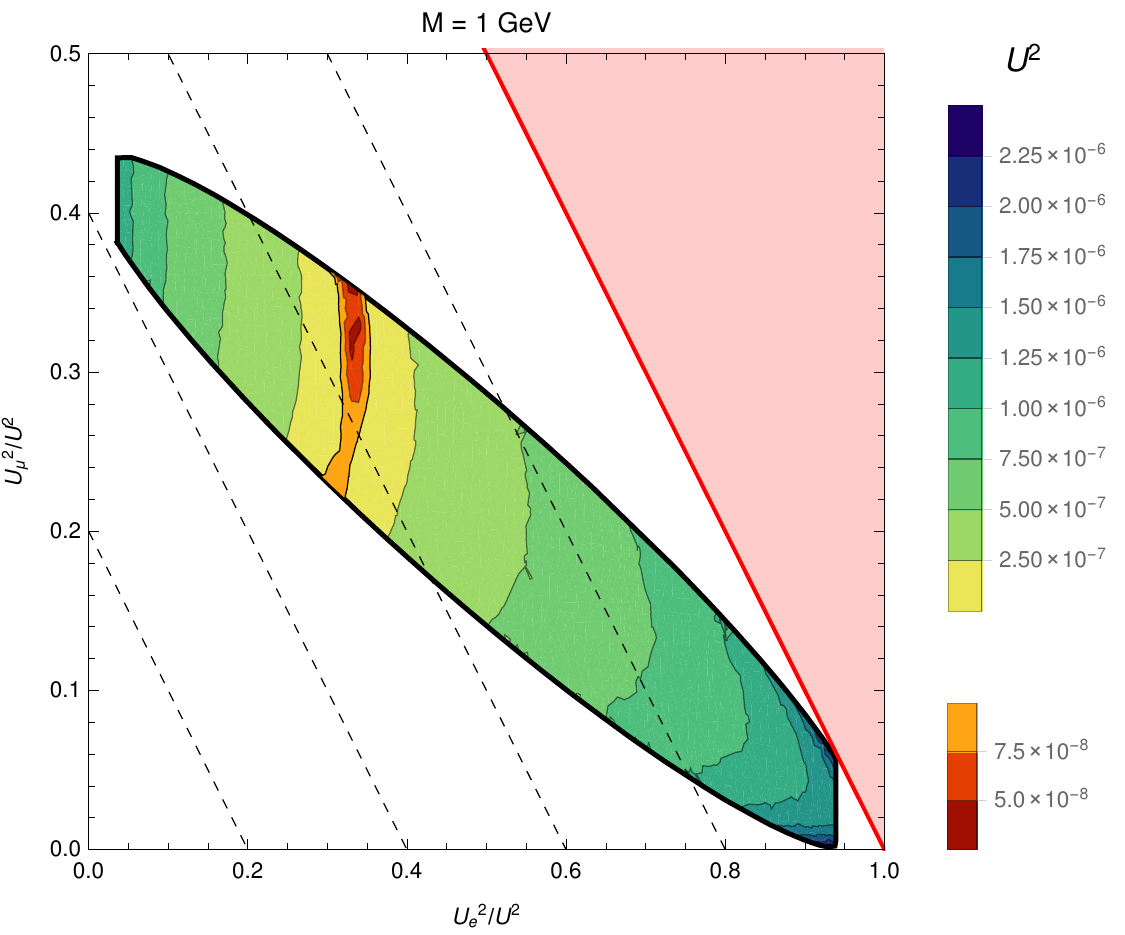}
	\caption{Constraints on $U_a^2/U^2$ from neutrino oscillation data, where $U_a^2=\sum_i U_{ai}^2$ and $U^2=\sum_a U_a^2$ \cite{Drewes:2016jae}.
Values inside the black line are consistent with neutrino oscillation data for normal hierarchy (left) and inverted hierarchy (right) of light neutrino masses. The dashed lines correspond to constant $U_\tau^2$, the light red region is unphysical, as it would require $U_\tau^2<0$.  The coloured regions indicate the maximally allowed $U^2$ for given $U_a^2/U^2$ if one requires that the observed $\eta_B$ can be generated by leptogenesis with $\bar{M}=1$ GeV. 
	\label{Fig:Lepto_NO}}
\end{figure}

\paragraph{Low scale leptogenesis} - 
The experimental perspectives for a discovery of the $N_i$ are most promising if they have masses $M_i$ below that of the W boson.
A discussion of the experimental constraints and perspectives can e.g.\ be found in the reviews 
\cite{Atre:2009rg,Ruchayskiy:2011aa,Drewes:2016jae,Kusenko:2009up,Ruchayskiy:2011aa,Boyarsky:2009ix,Drewes:2013gca,Alekhin:2015byh,Drewes:2015iva,Deppisch:2015qwa,Adhikari:2016bei}
 and references therein.
In this mass range, leptogenesis in the minimal model (\ref{eq:Lagrangian}) can only be achieved in the freeze in scenario.\footnote{
Note that this scenario may also work for larger masses \cite{Garbrecht:2014bfa}, so that there is an overlap between the $M_i$ ranges where the freeze out and freeze in scenario are feasible.
} 
Since the generation of the BAU happens at $T\gg M_i$, relies on flavour effects and may continue across the electroweak transition, all effects discussed above are potentially relevant in this scenario.
The viable parameter space has been studied in the models with two 
\cite{Canetti:2010aw,Canetti:2012vf,Canetti:2012kh,Shuve:2014zua,Abada:2015rta,Hernandez:2015wna,Drewes:2016lqo,Drewes:2016gmt,Hernandez:2016kel,Drewes:2016jae}
and three 
\cite{Drewes:2012ma,Shuve:2014zua,Canetti:2014dka,Hernandez:2015wna,Drewes:2016lqo}
heavy neutrinos.
A convenient way to illustrate the results is given by the projection in the $\bar{M}-U_a^2$ 
plane, where $U_a^2=\sum_i |\theta_{ai}|^2$. 
As an example, we show the $\bar{M}-U_\mu^2$ in Fig.~\ref{plot}.
If any heavy neutral leptons are discovered in the future,
an independent measurement of all $U_{ai}^2$, which could be done at the SHiP experiment or a future lepton collider, would provide a powerful test to asses whether these particles are the origin of neutrino masses and matter in the universe in the minimal seesaw model, see Fig.~\ref{Fig:Lepto_NO}.
Together with an observation of neutrinoless double $\beta$ decay and $CP$ violation in light neutrino oscillations, this would impose significant constraints on all model parameters \cite{Drewes:2016lqo,Hernandez:2016kel,Asaka:2016zib,Drewes:2016jae}.

\end{document}